# Structured Light at the Extreme

## Harnessing Spatiotemporal Control for High-Field Laser-Matter Interactions

**Abstract:** This review charts the emerging paradigm of intelligent structured light for high-field laser-matter interactions, where the precise spatiotemporal and vectorial control of light is a critical degree of freedom. We outline a transformative framework built upon three synergistic pillars. First, we survey the advanced electromagnetic toolkit, moving beyond conventional spatial light modulators to include robust static optics and the promising frontier of plasma light modulators. Second, we detail the optimization engine for this high-dimensional design space, focusing on physics-informed digital twins and AI-driven inverse design to automate the discovery of optimal light structures. Finally, we explore the groundbreaking applications enabled by this integrated approach, including programmable electron beams, orbital-angular-momentum-carrying γ-rays, compact THz accelerators, and robust communications. The path forward necessitates overcoming grand challenges in material science, real-time adaptive control at MHz rates, and the extension of these principles to the quantum realm. This review serves as a call to action for a coordinated, interdisciplinary effort to command, rather than merely observe, light-matter interactions at the extreme.

**Authors:** *Sergio Carbajo*[1-2*], *Seung-Whan Bahk*[3], *Justin Baker*[1], *Andrea Bertozzi*[1], *Abhimanyu Borthakur*[1], *Antonino Di Piazza*[3], *Andrew Forbes*[4], *Spencer Gessner*[2], *Jack Hirschman*[2], *Maciej Lewenstein*[5], *Yuhang Li*[1], *Inhyuk Nam*[6], *Eileen Otte*[3], *James Rozensweig*[1], *Yijie Shen*[7], *Liwei Song*[12], *Ye Tian*[8], *Yu Wang*[7], *Yuntian Wang*[1], *Logan Wright*[9], *Xiaojun Wu*[10], *Hao Zhang*[1-2]

[1]*University of California, Los Angeles, USA*

[2]*SLAC National Accelerator Laboratory and Stanford University, USA*

[3]*University of Rochester, USA*

[4]*University of the Witwatersrand, South Africa*

[5]*Institut de Ciències Fotòniques (ICFO), Spain*

[6]*Ulsan National Institute of Science and Technology, South Korea*

[7]*Nanyang Technological University, Singapore*

[8]*Shanghai Institute of Optics and Fine Mechanics, China*

[9]*Yale University, USA*

[10]*Beihang University, China*

*corresponding author: scarbajo@ucla.edu

*For review purposes only:*

## 1. INTRODUCTION

The ability to structure light—to tailor its amplitude, phase, polarization, and orbital angular momentum (OAM) in space and time—has fundamentally transformed optical science and technology over the past three decades[1,2]. Originally confined to the paraxial regime, pioneering work in singular optics and vector beams demonstrated that light's complex spatial structure could be exploited for applications ranging from optical trapping and micromanipulation to high-bandwidth communications, quantum electrodynamics, and quantum information[3–5]. This first wave of research has primarily dealt with static or slowly varying fields at low intensities, where devices such as spatial light modulators (SLMs) could easily shape the transverse electric field components of a laser pulse.

However, many of the most profound applications of structured light lie in the realm of high-field, high-intensity laser-matter interactions, where the electromagnetic (EM) field is strong enough to drive nonlinear, relativistic, and even quantum electrodynamic (QED) processes. In these regimes, the precise spatiotemporal structure of the laser pulse is not merely an accessory but a critical degree of freedom that can qualitatively alter the physics. For instance, it has been shown that OAM in laser pulses can be transferred to electrons, generating twisted γ-ray beams [6–8]—a potential gateway for novel nuclear photonics and secure space communications. Complex nonlinear Compton scattering effects are enriched with OAM phase transfer[9,10], and the use of vector beams can enhance electron trapping and acceleration to channel intense, coherent secondary radiation[11–13]. These discoveries represent just the initial forays into the vast and largely uncharted landscape of structured light at extreme intensities[14].

A critical challenge now impedes progress: with a limited understanding of which specific laser field structures are optimal, how can we systematically explore this high-dimensional and multi-dimensional parameter space to unlock new capabilities? Traditional, brute-force approaches, such as trial-and-error, are prohibitively inefficient and often fail due to the multi-scale, nonlinear complexity of these interactions. The current framework for light shaping is inadequate for the demands of high-power applications; conventional SLMs are unsuitable for shorter, ultraviolet (UV), or longer, mid- and long-wavelength infrared (MIR, LWIR), wavelengths and cannot withstand the extreme intensities required to probe new physics[15,16]. This gap between the need for exquisitely tailored light and our ability to create it represents a fundamental bottleneck.

Bridging this gap necessitates a paradigm shift from conventional beam shaping to a new era of intelligent structured light at the extreme, which unites three foundational pillars: (1) advanced optical techniques for generating high-intensity, reconfigurable spatiotemporal vectorial EM fields; (2) sophisticated diagnostics to characterize these complex fields; and (3) artificial intelligence (AI) and inverse design to automate the discovery and optimization of light-matter interactions. This confluence is poised to revolutionize our command over particle acceleration, plasma dynamics, and secondary radiation generation, moving the field from using unstructured light "hammers" to employing a full set of intricately tailored, high-intensity light tools (see synergistic strategy in Fig. 1).

This review article charts the course for this emerging field. We begin by surveying the state-of-the-art electromagnetic toolkit (Sec. 2), exploring innovations beyond SLMs and other shaping devices, including robust static optics like axicons and diffractive gratings, and the promising new frontier of plasma light modulators (PLMs) for programmable control at any wavelength and intensity. We detail the journey towards full vectorial control of all electric $E = [E_x, E_y, E_z]^T(x, y, z, t)$ and magnetic $H = [H_x, H_y, H_z]^T(x, y, z, t)$ field components in 4D space-time.

We then delve into the optimization engine (Sec. 3) required to navigate this complexity. We discuss the inverse problem in multi-scale physics and present a transformative approach: the development of modular, physics-informed digital twins of entire experiments. Frameworks like physics-aware training (PAT) enable end-to-end gradient-based optimization via backpropagation, allowing researchers not only to find an optimal pulse but also to understand why it works[17,18]. Furthermore, automated novelty search algorithms can identify entirely new archetypes of light-matter interaction regimes.

Finally, we explore the transformative frontiers in application (Sec. 4) that such tailored light enables, from advanced X-ray free-electron laser (XFEL) modalities and the generation of OAM γ-rays for nuclear photonics to laser-based collimation for future particle colliders and structured THz beams for communications. Progress in this field relies on a synergistic strategy that brings together experts in laser physics, accelerator science, plasma physics, and machine learning.

The path forward is exceptionally challenging, requiring new materials, real-time adaptive control systems, and integrated co-design of optics and algorithms. This review aims to synthesize the current state of the art, identify the most pressing grand challenges, and serve as a call to action for a community poised to usher in a new revolution in light-matter interactions.

## 2. THE ELECTROMAGNETIC TOOLKIT: Generating Structured Light for High-Power Applications

The controlled application of structured light to high-field physics demands a new generation of optical tools[19], but existing high-power solutions based on external modulation are largely static in nature[20]. Generating high power and high intensity structured light directly at the source is very much in its infancy, with the toolkit largely restricted to preconfigured outputs. On-demand control outside the source requires moving beyond the capabilities of conventional SLMs: this new toolkit must handle high peak and average powers, operate across a broad spectral range from UV to THz, and provide reconfigurable control over the complete spatiotemporal and vectorial state of an electromagnetic field. This section surveys the critical components of this toolkit, from robust static elements to dynamic programmable devices and the advanced diagnostics required to characterize them.

### 2.1. High-intensity structured light at the source: Challenges and opportunities

It is often convenient to create structured light directly at the source, but despite the tremendous progress in structured light lasers over the past decade[21–24], the advance towards high-power and high-intensity has been somewhat limited[25]. Traditional solid state laser solutions with either intra-cavity or gain control have typically produced structured light of average power in the <100 W range, limited by the power handling capability of the shaping devices themselves (Section 2.2), exacerbated by the intracavity enhancement factor. High-power reflective shaping elements have fared much better, allowing gas and disk lasers to reach multi-kilowatt powers[26] or relativistic intensities[27,28]. Average powers in both scalar and vectorial light can be enhanced by amplification, either in bulk or fibre-based systems, with the main limitation the balancing of power extraction and modal purity[29,30]. An interesting approach that is gaining traction is coherent beam combining as a route to high-power structured light[31–33]. Here, a seed laser is split into multiple channels, each amplified and recombined. Dynamic control of the initial beams means that by coherent addition, the output is high-power (> 100 kW) yet dynamic in output structure[34]. Peak powers and intensities can likewise be achieved if the pulse duration is made short, usually by mode locking. Amplification of shaped light from the source has seen ~GW peak powers of vectorial light after multiple passes[35].

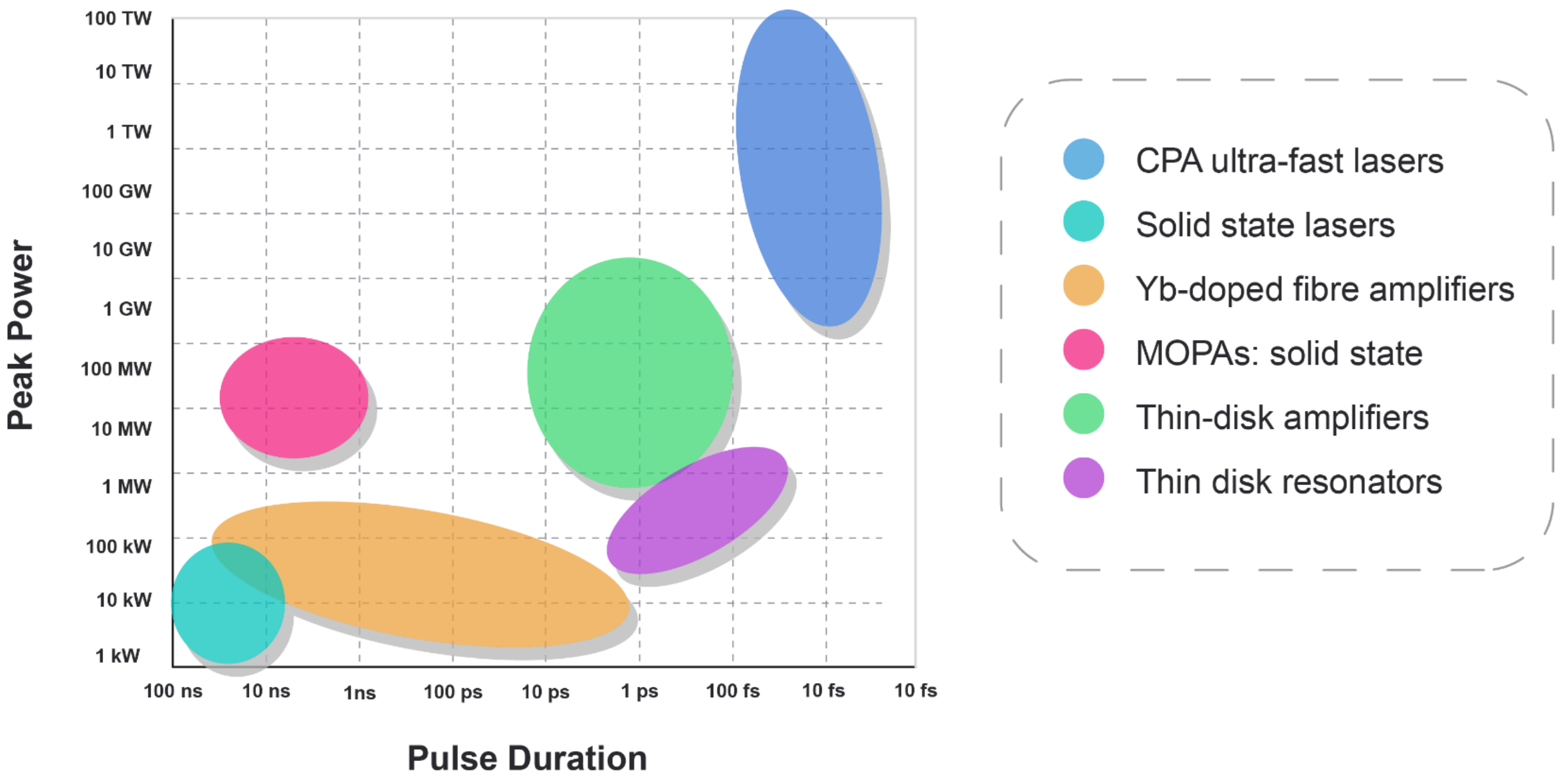

**Fig. 1 | Overview of strucutred light source brightness.** Methods of structured light generation by source as a function of experimentally demonstrated peak power and pulse duration.

Despite the advances, high-power and/or high-intensity structured light lasers have largely been limited to highly sophisticated laboratories, yet a versatile toolkit accessible to all is required for widespread deployment of the technology.

### 2.2. Beyond Spatial Light Modulators: The Challenge of High Intensity and Broad Bandwidth

Although conventional devices for shaping light[36], such as acoustic-optical modulator (AOM)[37,38], acousto-optic programmable dispersive filters (AOPDF)[39,40], liquid-crystal-based SLMs[41,42], and digital micromirror devices (DMDs)[43,44], are able to manipulate ultrafast light, they are nevertheless largely restricted to low average powered structured light. They operate by imparting a spatially dependent phase and/or amplitude modulation on a wavefront, enabling the generation of modes carrying OAM and other complex scalar fields[1,2]. However, their utility diminishes drastically in high-power regimes. They are susceptible to optical damage[45], exhibit limited efficiency, and are often restricted to a narrow band of the spectrum, typically the visible and near-infrared[15]. This makes them unsuitable for the shorter (UV) and longer (MIR, LWIR) wavelengths that are optimal for many advanced light-matter interactions, such as driving photocathodes or exciting molecular vibrations.

The central challenge is thus to develop shaping technologies that can withstand high intensities and operate across a wide bandwidth without compromising on the dimensionality or fidelity of control. While recent commercial platforms are now rated for >1 kW CW operation[46], marking a major step toward practical high-fluence beam shaping, there are still limitations in pulse energy and wavelength range. This has driven innovation towards two complementary paths: the use of robust, passive static optics designed for specific transformations and the development of novel programmable plasma-based modulators that circumvent the damage thresholds of solid materials.

### 2.3. Static Optical Elements: Diffractive Gratings, Axicons, and Free-Form Optics for Robust Beam Shaping

For many applications, a reconfigurable device is unnecessary; a precise, static transformation is sufficient. In high-power laser systems, static elements like diffractive gratings (DGs) and axicons can be the key. Axicons, for example, are conical lenses that transform a Gaussian beam into a propagation-invariant Bessel beam, a property exploited to generate extended plasma channels for wakefield acceleration[47] and advanced X-ray production regimes[48,49]. These components are typically manufactured from durable substrates like fused silica and can be coated to achieve high damage thresholds, making them ideal for intense laser pulses.

Recent advancements focus on extending the functionality of these static components. Researchers are designing polarization-insensitive and polarization-sensitive diffractive optics. This includes not only lenses but also custom gratings that may perform a change-of-basis transformation, for instance, converting a standard Gaussian beam into a Bessel-type field or directly implementing a Fourier-space modulation on a ring-shaped subspace[50,51]. Furthermore, the use of birefringent crystals like Potassium Dihydrogen Phosphate (KDP) enables the fabrication of free-form optics for vectorial control [52,53]. Spiral-shaped KDP elements could act as high-power vortex waveplates (q-plates [52,53]) to generate cylindrical vector beams, which are crucial for creating focal fields with strong longitudinal electric or magnetic components[54] for advanced electron-beam control (see example focal field in Fig.

2.f, bottom). Note that these optics can perform real space vectorial modulation (Fig. 2.f, (i)), while also embedding wavelength-dependent properties – similar to the realization of a “flying focus”, taking advantage of chromatic effects of diffractive grating[55]. Alternatively, the static optical elements can be combined with established 4f pulse shaping (Fig. 2.f, (ii)) for direct spectral vectorial control.

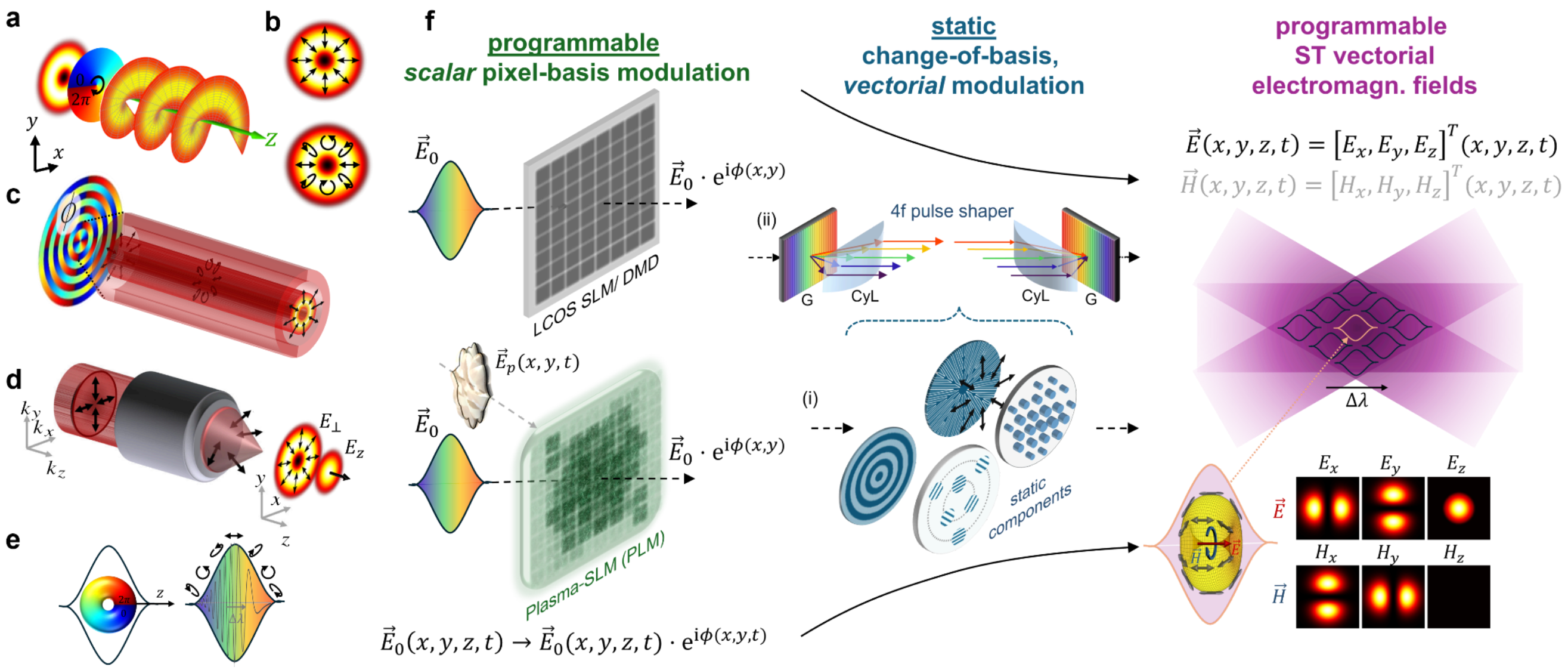

**Fig. 2 | Reshaping the framework of ST modulation.** (a-e) Past development of structured light from paraxial (a) scalar (OAM) and (b) vector fields shaped in 2D space, to (c) their 3D spatial modulation, as well as (d) tight focusing into the non-paraxial regime with 3D oscillating components, and (e) ST modulation. (f) Customized ST vectorial EM fields facilitated by combining reconfigurably shaping scalar phase and amplitude by an SLM/DMD (low power/intensity) or PLM (high power/intensity) with vectorial modulation by the use of static optics (in (i) real space or (ii) spectrally); this allows to transform from the “pixel” basis of scalar modes to non-paraxial, vectorial modes, e.g. in Laguerre-Gaussian or Bessel basis.

### 2.4. Dynamic and Programmable Scalar Control: The Promise of Plasma Light Modulators (PLMs)

To achieve true reconfigurability at high intensities, the most promising approach is to use light to control light via a plasma-based medium. Plasma Light Modulators (PLMs) represent a paradigm shift. The concept involves using a precisely shaped, low-energy "pump" laser pulse to ionize a medium, initially a solid target like silicon-on-insulator for experimental simplicity, later transitioning to gases or liquids, to create a tailored spatiotemporal pattern of electron density, ρ(x,y,z,t)[56,57]. A subsequent, high-intensity "probe" pulse propagating through this plasma structure experiences programmable phase and amplitude modulations, effectively being shaped by the plasma's transient refractive index profile.

The critical advantage of PLMs is their ability to operate at essentially any wavelength and at intensities that would instantly destroy conventional solid-state optics. Their development can uniquely be enabled by the inverse design procedures discussed in Section 3[17] [13]. Instead of trying to manually design a complex plasma structure, algorithms can optimize the pump laser field $E_p(x, y, t)$ such that the generated plasma produces a user-defined effect on the probe pulse. Early experimental milestones could target the demonstration of a scalar PLM capable of realizing arbitrary spatial phase modulations with at least ~1000 effective "pixels"[58], paving the way for future portable modules that can be integrated into major petawatt-class and beyond light sources.

### 2.5. The Full Vectorial Frontier: Crafting Full Electromagnetic Fields with 3D Polarization Control

True mastery of light-matter interactions requires control beyond scalar phase and amplitude; it demands command over the full vectorial nature of the electromagnetic field. In general, a light field is described by its three-dimensional electric and magnetic vector components $E = [E_x, E_y, E_z]^T(x, y, z, t)$, $H = [H_x, H_y, H_z]^T(x, y, z, t)$, which vary in 4D space-time. This is particularly critical in high-intensity interactions, where the polarization state plays a crucial role in controlling plasma dynamics and acceleration processes[59].

A powerful strategy to access this vectorial frontier is the use of an optimal modulation basis. Bessel modes are a particularly attractive choice[60–62]. Unlike standard Laguerre-Gaussian modes, Bessel beams are exact, non-paraxial solutions to the wave equation that naturally include all three electric field components. They form a complete orthonormal set, allowing access to arbitrary spatiotemporally structured vectorial light fields. Their characteristic ring-shaped Fourier-space distribution also simplifies experimental complexity: complex structures can be achieved by modulating amplitude, phase, and polarization on simple ring-shaped subspaces in Fourier space, which are then transformed to real space by a static optic through a Fourier transforming lens (see spatial modulation examples in Fig. 3).

By combining the programmable scalar control of a PLM (or, for lower intensity, SLM) with static change-of-basis optics (e.g., axicons) and vectorial control components (e.g., KDP optics), researchers can now generate previously inaccessible fields (cf. Fig. 2.f). These will include new types of flying foci and spatiotemporal optical vortices (STOVs), non-diffracting arrays of counter-rotating OAM beams, and other light fields with strong longitudinal modulation in ST [63,64]. This comprehensive approach includes the generation of focal fields that mimic the structure of electric or magnetic dipoles ("toroidal pulses") [[63,64]] and multipoles, opening new doors for controlling charged particle dynamics and light-matter interactions at the extremes.

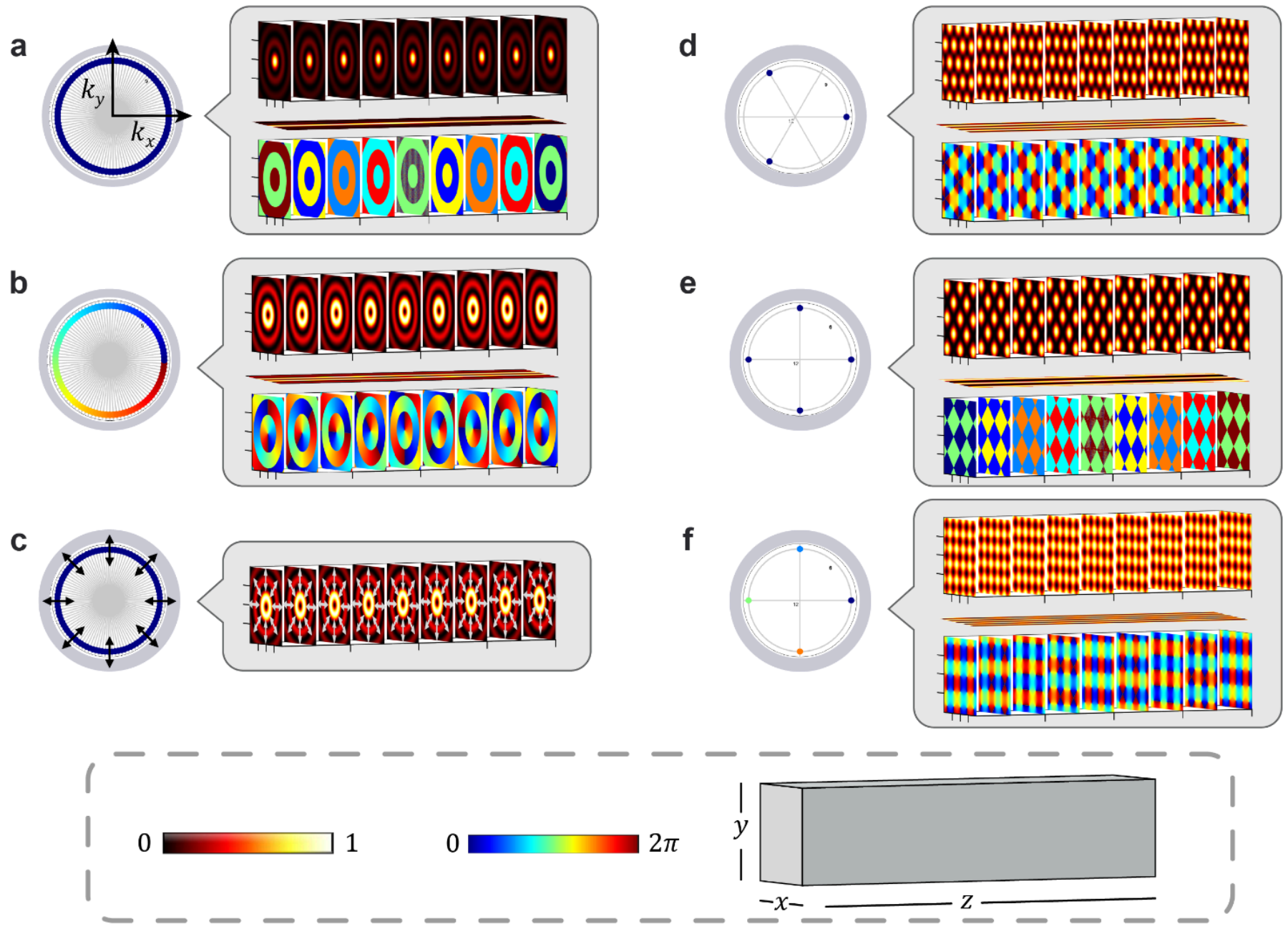

**Fig. 3 |** The example of the Bessel basis for smart modulation (left, right: Fourier/k-, real space; single frequency). Programmable customization is performed by Fourier modulation (SLM/DMD, PLM, or apertures) plus advanced static (vectorial) optics and transformation to real space.

### 2.6. Critical Enablers: Advanced Spatiotemporal Diagnostics for Single-Shot Characterization

The creation of these complex fields is only half the challenge; they must also be measured. Traditional diagnostics are inadequate for characterizing the complete spatiotemporal and vectorial state of a high-intensity pulse. This has spurred the development of advanced single-shot, multispectral diagnostics.

Techniques such as single-shot multispectral wavefront sensing [65] provide a detailed picture of a pulse's spatio-spectral properties. These can be further enhanced using meta-optics and nonlinear optics to enable full-time-domain measurements. For instance, single-shot cross-correlation techniques using random quasi-phase-matching crystals [66] can be used to map a complex spatiotemporal pattern in the focal region onto an imaging device. These diagnostics are essential for closing the loop in adaptive control systems. A baseline system can use a simple near-field objective to provide feedback, but a more powerful approach would directly incorporate measured far-field spatiotemporal properties into the optimization loop. This requires diagnostics that can provide physics-based parameters as feedback, enabling real-time optimization of phenomena like velocity-matched "flying focus" pulses that can travel at any velocity, including superluminally, for applications in photon acceleration and remote sensing[67,68].

In conclusion, the electromagnetic toolkit for high-power applications is rapidly evolving from a collection of simple static elements to an integrated suite of static, programmable, and diagnostic technologies. This toolkit provides the essential foundation for the inverse design and optimization of light-matter interactions, which we will explore in the next section.

## 3. THE OPTIMIZATION ENGINE: Inverse Design and AI for Light-Matter Control

The staggering complexity of high-intensity light-matter interactions, often spanning multiple coupled spatial and temporal scales and involving strong nonlinearities, renders traditional design methodologies based on human intuition obsolete. Simply knowing that a shaped light field can alter an interaction is insufficient; the central challenge is to discover which specific, high-dimensional spatiotemporal structure will optimally drive a system toward a desired outcome. The problem of optimally designing laser fields directly for tailored light-matter interactions has existed for decades and was studied extensively in the context of quantum coherent control, e.g., of chemical reactions or molecules[69–73]. However, modern advances in computation, machine learning and high-dimensional programmable shaping of high-intensity light fields have renewed interest in this challenge, particularly in the more expansive context of spatiotemporal and vectorial control of laser fields[16,74–77], and beyond the setting of quantum coherent control of chemistry that originally inspired the field. Although the challenge of on-demand, task-specific spatiotemporally tailored laser light remains unsolved, this section details the emerging paradigm that aims to meet this opportunity, an approach broadly characterized by coupled physics-based and machine learning (ML) models, and adaptive experimentation to create an automated pipeline for the inverse design of light-matter physics. This challenge will require deep collaborations between researchers in domains of physics-informed machine learning[78,79], alongside theorists and experimentalists in laser-mediated science.

### 3.1. The Inverse Problem: Why Trial-and-Error Fails in Multi-Scale Physics

The inverse problem—calculating (designing) the input laser field required to produce a specific experimental output—is notoriously difficult in high-field physics. The laser-driven experiments and high-intensity laser sources themselves are usually inherently nonlinear and involve essential coupling between distinct physical processes across a range of scales (e.g., from atoms and x-rays to infrared photons and collective phases of matter) that are difficult for even experienced human scientists to reason about. Changing input laser parameters (e.g., spatial and temporal phase, polarization) can lead to disproportionate and counterintuitive changes in the outcome (e.g., electron beam emittance, γ-ray spectrum, plasma density). Furthermore, first-principles simulations, while invaluable, are frequently too computationally expensive for broad parameter searches and suffer from the "simulation-to-reality" (sim2real) gap due to unknown experimental parameters and unmodeled physical effects.

Consequently, a brute-force, trial-and-error approach is profoundly inefficient and unlikely to discover truly optimal or novel solutions within a vast, high-dimensional parameter space. A more effective route is a data-driven, adaptive learning loop that augments physics models or directly closes the loop on the experiment—e.g., Bayesian optimization for laser–plasma accelerators and other laser systems[77,80–82], deep reinforcement-learning control for coherent beam/pulse shaping[83,84], and adjoint/ inverse-design methods with differentiable solvers or surrogates for photonics and ultrafast optics[77,85–89]. Together these approaches efficiently navigate complex control spaces while remaining compatible with experimental feedback, offering a practical path to solve the inverse problem.

To meet the challenge of real-world inverse design of laser-driven processes, the field will need to combine innovations across three distinct domains: (1) Highly reconfigurable, high-throughput experimental apparatuses for shaping intense laser light (as described in Section 2), as well as (2) digital twins to accurately, differentiably model laser-driven experiments, and (3), optimization and search algorithms that leverage (1) and (2) to optimize or discover user-defined targets in the laboratory.

### 3.2. Digital Twins: Building Physics-Informed and Machine-Learned Models of Complex Experiments

A powerful strategy for managing the complexity of sophisticated experiments is the construction of a modular digital twin (DT) of the entire experimental setup. A DT is a computational model that mirrors the behavior of a physical system. In the context of a complex laser-driven experiment, conceptualized in Fig. 4, the final measurement M (e.g., X-ray spectrum) depends on a long sequence of processes: the initial laser pulse shaping, nonlinear propagation, interaction with a plasma or electron beam, and radiation generation.

A modular DT describes this chain as a composition of functions: $M = f(g(h(\theta)))$, where $\theta$ represents the control parameters of the initial laser field. The key insight is that different stages can be modeled with varying degrees of fidelity. Some stages are well-described by established physics codes (e.g., particle-in-cell simulations for plasma interactions, FEL codes for radiation generation), while others, too complex for simple models, can be represented by machine-learned models, such as neural networks (NNs), trained on experimental data[18,90].

Frameworks like the Lightsource Unified Modeling Environment (LUME) are being developed to integrate these disparate simulation codes into a unified, user-friendly ecosystem[91]. This modular structure is not just for prediction; it is designed for optimization. By making the DT differentiable, one can use autodifferentiation and backpropagation—the engine of modern deep learning—to calculate how a change in the initial laser parameters $\theta$ will affect the outcome $M$. This allows researchers to efficiently compute the gradient of a loss function $\mathcal{L}(\hat{M}, M)$ with predicted and desired measurements $\hat{M}$ and $M$, respectively. The result is precise optimization of the laser field to achieve a target experimental result, just as one would train the weights of a NN[17].

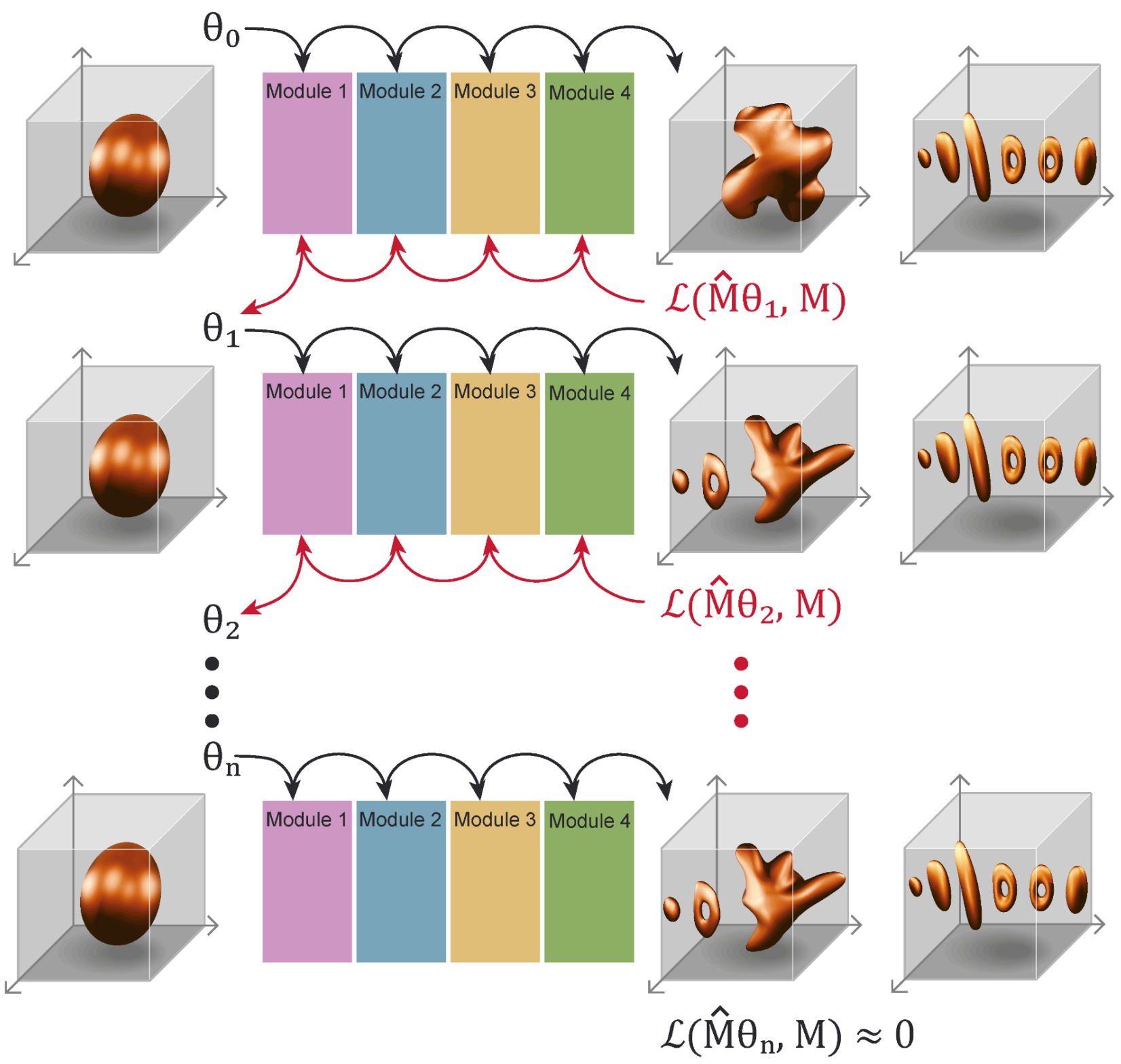

**Fig. 4 |** Diagramatic physical and DT representation of the LCLS-II XFEL, from the photoinjector laser system all the way to attosecond time-resolved shot-to-shot X-ray diagnostics.

The use of digital twins, whether modular or not, as a basis for computational optimization is a trend encompassing a wide range of optical experiments. For example, a similar principle underlies the inverse design of optical systems like deep diffractive neural networks ($D^2$NNs), where free-space propagation between diffractive layers is modeled by scalar diffraction theory and the transmission profiles of each transmissive layer are optimized, at the diffraction limit of light, to achieve target optical responses[92]. Such architectures effectively act as differentiable digital twins of the diffractive optical processor, combining wave optics-based forward models with data-driven statistical learning. Likewise, convolutional neural networks have been used to approximate free-space holographic propagation, learning the mapping governed by diffraction integrals and enabling fast surrogate modeling of complex optical fields[93,94].

### 3.3. Optimizing and autonomously exploring laser-driven processes in situ

With or without a digital twin for an experimental laser-driven process, researchers may deploy a variety of algorithms aimed at optimizing a physical objective function, such as the amount of power of a secondary radiation produced, by choice of the freely adjustable experimental controls. Currently, there are several choices available to researchers seeking to optimize such a reconfigurable experiment[95].

In absence of a digital twin, black box heuristic optimization algorithms such as genetic algorithms (GA), model-free reinforcement learning, and parameter exploring policy gradients (PEPG) can be deployed[96–100]. These algorithms require virtually no user knowledge beyond a well-defined objective function and are thus straightforward to apply. However, the scaling behavior of these algorithms with respect to the number of optimized parameters is typically much worse than algorithms that can exploit more detailed knowledge of the underlying physical process, such as that provided by a digital twin.

The availability of a digital twin, even an approximate or fully data-driven one, can allow potentially more powerful optimization algorithms to be deployed, albeit typically with the cost of more difficult implementations. Ideally, the digital twin, a differentiable digital representation of the map between experimental control parameters and experimental measurements, can be used as the basis for a gradient-based optimization, employing methods like autodifferentiation and the adjoint method. Methods in this family, often referred to as differential programming methods, are closely related to the now famous backpropagation algorithm that has become the ubiquitous basis for training of modern large-scale networks, owing to its exceptional scaling behavior. Once the parameters of the digital twin are optimized, they can then be transferred to the experimental apparatus. While there are some systems, like the diffractive networks mentioned above, where this simulation to reality transfer can be realized with minimal performance loss, in most highly nonlinear laser-driven processes, small deviations between the experiment and its digital twin can translate to enormous performance drops – a harsh reality known in the robotics literature as the simulation-reality gap[17,101,102].

To overcome the simulation-reality gap, a number of emerging methods have been proposed, though their application to control of complex, nonlinear laser-driven experiments remains untested. For example, physics-aware training (PAT) uses a digital twin to efficiently compute parameter updates using backpropagation, but makes use of the physical process itself on the forward pass. If the digital twin is sufficiently accurate, this hybrid “experiment in the loop” approach may dramatically suppress the effects of simulation-reality gap[17,103]. However, PAT is just one possible approach among many conceivable hybrid algorithms for laser-driven experiments that seek to combine the scaling of gradient-based optimization with the robustness and ease of use of black-box heuristic optimizers. For example, a growing number of works, spanning an impressive range of domains, have shown that carefully imposing a variety of noise and imperfections in simulations during training can result in simulation-derived optimal parameters can be transferred accurately to experiments[104,105].

Finally, while optimization algorithms like those described above offer a solution when a well-defined objective can be specified, they are not necessarily suitable for open-ended discovery, which is often the objective of extreme intensity laser-driven experiments. While this open-ended search for interesting physical phenomena shares many commonalities with objective-driven optimization, particularly when the search is intended to seek out phenomena with user-specified characteristics, it also involves new challenges, such as specifying an objective that captures what human researchers consider “intersting”. Here, intelligent search algorithms, such as those utilized in reinforcement learning, will be necessary[106,107]. Alternatively, given access to code-controlled experiments, AI agents based on large language models could be used in place of human explorers, seeking out experimental regimes that could even be specified qualitatively by a managing human scientist[108,109].

In summary, the integration of programmable, complex laser field shaping, modular digital twins, and automated optimization and discovery algorithms forms a powerful engine for high-dimensional exploration and optimization of laser-driven experiments. This engine could one day soon transform the design process from one of manual intuition to an automated, data-driven search, capable of both achieving predefined goals with unprecedented efficiency and uncovering entirely new realms of physics hidden within the complexity of high-field interactions.

### 3.4. AI for light–matter inverse design

Modern optoelectronic devices exhibit strongly coupled, high-dimensional, and deeply subwavelength behavior, making classical intuition-driven design inadequate. While forward solvers can compute optical responses for known structures, the inverse problem—recovering geometry from desired functionality—is non-unique and analytically inaccessible. Machine learning enabled reverse design resolves this mismatch by transforming complex structure response relationships into learnable, differentiable mappings[110–113].

Discriminative neural networks provide fast surrogate models for metasurfaces, multilayer films, and resonant particles, enabling gradient-based optimization far beyond the reach of manual parameter sweeps[114–116]. Joint forward inverse architectures successfully retrieve metasurface geometries for polarization-dependent spectra, while surrogate-assisted backpropagation optimizes multilayer nanostructures and freeform photonic crystal cavities, achieving ultrahigh-Q designs without adjoint derivation[117–119]. Yet as the design space expands into thousands of degrees of freedom, discriminative models alone become insufficient, motivating the shift toward generative approaches that restructure the geometry space itself. Generative models such as GANs and VAEs address the curse of dimensionality by learning compact latent manifolds of physically realizable geometries, allowing efficient optimization over thousands of design variables[120,121]. Hybrid workflows—GAN generation plus adjoint refinement, or generative–evolutionary methods like GLOnet combine global exploration with local optimality and routinely converge to near-global optima in complex design spaces[122,123].

The emergence of universal photonic surrogate models—capable of predicting full 3D polarization density and associated far-field, thermal, and nonlinear responses—further elevates reverse design into a multi-objective optimization framework. These models link physical law with data-driven inference, allowing simultaneous optimization across performance metrics essential for modern LEDs, detectors, nonlinear converters, and integrated photonics [124,125].

Beyond device-level geometry optimization, the principles of inverse design naturally extend to optical field reconstruction itself. A parallel inversion paradigm emerges in structured-light systems. When topological beams traverse disordered media, their amplitude, phase, and polarization information becomes experimentally inaccessible; a topology-enhanced deep-learning framework can nevertheless recover intrinsic optical topology from single-shot speckle patterns with ~92% accuracy, enabling secure structured-light communication. This mirrors the core principle of ML-based inverse design: learning reconstructs high-dimensional optical information that traditional methods cannot access[126,127].

Together, these advances establish reverse design as a foundational methodology for next-generation optoelectronics. Machine intelligence reshapes the design landscape, enabling freeform geometries, topology reconstruction, and multifunctional operation far beyond the limits of conventional photonic design strategies.

## 4. FRONTIERS IN APPLICATIONS: Breakthroughs Enabled by Tailored Light

The convergence of advanced light shaping, AI-driven optimization, and high-power laser technology is not an end in itself; its true value is realized in the transformative applications it enables. By providing unprecedented control over the initial conditions of extreme light-matter interactions, this integrated approach is poised to revolutionize fields from fundamental physics to national security. This section details the groundbreaking advancements currently underway across four key domains.

### 4.1. Next-Generation Light Sources: Programmable Electron Beams and Advanced XFEL Modalities

The performance of linear accelerators and XFELs is fundamentally constrained by the quality of the electron beam generated at the photocathode. A primary figure of merit—beam brightness—is governed by the six-dimensional phase-space volume of the emitted bunch[128–133], and directly determines the X-ray beam brightness, which quantifies the photon flux per unit source area, solid angle, and bandwidth. Spatiotemporally shaping the ultraviolet (UV) photocathode-driving laser pulse is a powerful method to tailor this electron phase space at birth, offering a direct path to brighter, higher-current beams[134–139]. Early experiments demonstrated that tailored spatial and temporal laser profiles can measurably reduce thermal emittance and enhance beam laminarity, establishing the experimental foundation for laser-based phase-space engineering[134,136–140], as depicted in Fig. 5, including the first experimental demonstration of a mode-locked XFEL[141].

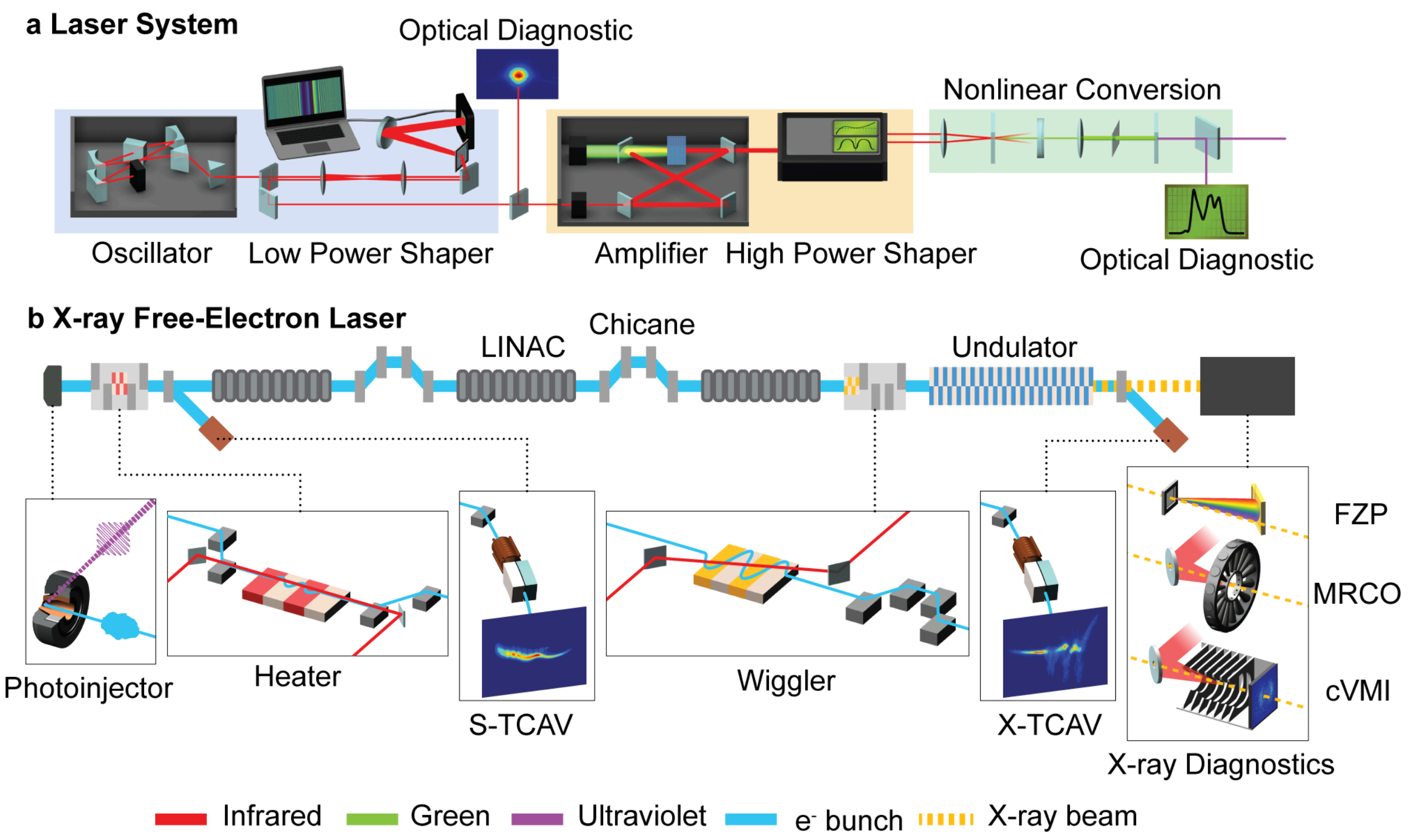

**Fig. 5 | Conceptual schematic of a programmable photoinjector laser and XFEL beamline.** a) Generalized architecture of a programmable photoinjector laser system. An oscillator and low-power shaping stage define the initial spatiotemporal field, which is subsequently amplified and further shaped using high-power spatial light modulators, plasma-based modulators, or static diffractive optics. Nonlinear upconversion delivers the ultraviolet photocathode-drive wavelength, with optical diagnostics monitoring performance throughout. b) Schematic of the major components in a modern XFEL: a photoinjector where the UV laser generates and shapes the electron bunch; a laser heater for microbunching control; linac and chicane sections for acceleration and compression; and dedicated modulator undulators (wigglers) for laser-driven energy modulation prior to the final X-ray–producing undulator. Downstream diagnostics—including S-band and X-band transverse deflecting cavities (TCAVs) and a suite of X-ray spectrometers and imaging detectors (FZP, MRCO, cVMI)—enable single-shot characterization of both electron and photon pulses. Together, these subsystems provide the foundation for advanced tuning and adaptive control strategies aimed at next-generation, software-defined light sources.

In practice, however, precise shaping in the UV remains technically formidable. Most photocathodes require excitation near 250–270 nm–such as copper or cesium telluride[142–145]–, where optical materials and liquid-crystal spatial light modulators (SLM) exhibit strong absorption and rapid degradation. As a result, shaping is typically performed earlier in the infrared or visible front end, where a broad suite of static and programmable devices—4f-line pulse shapers, acousto-optic modulators and dispersive filters, and liquid-crystal SLMs—can operate safely at modest fluence and large aperture. The target field is then propagated through subsequent amplifier and frequency-conversion stages. Because typical upconversion techniques introduce a nonlinear mapping between the designed IR field and the realized UV waveform, the final photocathode illumination often differs substantially from the programmed shape. This motivates the inverse-design and digital-twin frameworks introduced in Section 3 to predict and compensate for those transformations within full start-to-end beam-dynamics models.

Recent work has begun to bridge these wavelength and power limitations. Dispersion-Controlled Nonlinear Synthesis (DCNS) has demonstrated symmetric and asymmetric multi-μJ UV pulse generation at MHz repetition rates for the LCLS-II photoinjector [146], achieving shapes theoretically predicted to reduce emittance by ≈30 % and increase peak current by 2–3×[146,147]. Advances in high-damage-threshold reflective SLMs and digital micromirror devices (DMD) have extended programmable shaping to the post-amplifier IR regime[43], while Plasma Light Modulators (PLMs) introduced in Section 2 promise a transformative leap: plasma-based, transient refractive-index modulation that can imprint arbitrary amplitude and phase patterns on high-energy pulses at essentially any wavelength[56,57,148,149]. Unlike conventional solid-state devices, PLMs are intrinsically damage-immune and scalable, enabling dynamic shaping directly in high-fluence stages where optical materials would otherwise fail.

At the same time, a complementary materials effort aims to shift the operating wavelength of next-generation electron sources. Visible-wavelength photocathodes—for instance based on $CsK_2Sb$ or GaAs derivatives—can operate under green-to-blue excitation, alleviating UV-optics damage constraints and enabling direct programmable shaping with mature visible-light devices[150,151]. Combining such visible-band cathodes with PLM-enabled, post-amplifier control could allow continuous, damage-resilient operation across the full photoinjector system. Parallel advances in phase-preserving nonlinear upconversion, including recent four-wave-mixing (FWM) studies demonstrate high fidelity in phase transfer from IR to UV, offering an alternate strategy for faithful waveform replication beyond conventional harmonic generation[152,153].

Integrating these shaping technologies at the photoinjector with real-time electron beam and x-ray diagnostics—from transverse deflecting cavities (TCAVs) for the electron bunch to fresnel zone plate (FZP) spectrometers and angular streaking-based diagnostics like the Multi-Resolution COokiebox (MRCO) and co-linear velocity map imaging (cVMI)[154–164], creates a feedback network capable of adaptive optimization across the accelerator chain. Coupling these diagnostics to differentiable, physics-informed digital twins (Section 3.2) allows gradient-based optimization of the laser field parameters to manipulate target beam characteristics such as brightness and temporal shape[91,165–167].

Downstream of the injector, laser-based control continues through the laser heater, traditionally used to suppress microbunching instabilities that degrade beam quality and reduce FEL performance[48,168]. By resonantly overlapping an electron bunch with a laser pulse in a short undulator embedded within a magnetic chicane, the heater introduces a controlled slice energy spread proportional to the local laser field[169,170]. Beyond this stabilizing role, temporal or spectral shaping of the heater laser has emerged as a tool for direct phase-space control. Tailoring the laser envelope enables selective spoiling to generate femtosecond X-ray pulses, spectral modulations in seeded FELs, or current-profile perturbations that yield attosecond bursts and stabilize short-pulse emission in cavity-based systems [171–174]. Because the heater laser can be arbitrarily timed with respect to the electron bunch, it also enables pulse-by-pulse customization in high-repetition-rate operation—an experimental realization of beam-à-la-carte multiplexing across beamlines[49]. Coupling laser-heater shaping with upconversion techniques in the photoinjector laser, such as DCNS, may further enhance emittance control and X-ray brightness[131,135,146,147]. Looking ahead, periodic energy modulation via a tailored heater could seed mode-locked FEL operation, producing coherent X-ray frequency combs and trains of attosecond pulses[175]. Further downstream beyond the laser heater, prior to final X-ray production, a dedicated modulator undulator—often referred to as a wiggler—with a period matched to an externally or internally seeded IR laser can imprint a strong energy modulation on the electron beam. After a subsequent dispersive section performs the requisite phase-space rotation, this energy modulation converts into a sharp temporal density spike. Such pre-bunched, high–peak-current spikes radiate coherently in the final undulator, enabling the generation of isolated attosecond X-ray pulses [176–178].

These developments transform the XFEL from a fixed-configuration instrument into a software-defined, reconfigurable photon factory. By uniting low-power programmable shaping in the IR or visible, high-power laser modulation, specially selected nonlinear upconversion techniques, and data-driven feedback control, next-generation light sources will be able to program their emission characteristics, from pulse structure and spectrum to coherence and timing, in real time. This convergence of optical engineering, materials innovation, and digital-twin intelligence establishes the foundation for fully programmable electron beams and advanced XFEL modalities that adapt dynamically to the scientific demands of each experiment[165,166,179].

### 4.2. Structured Nuclear Photonics and Quantum Electrodynamics: γ-Rays and Orbital Angular Momentum Transfer

The advent of nuclear photonics has ushered in a transformative era for interdisciplinary science, enabling unprecedented control over nuclear processes with profound implications across physics, materials science, and medicine. This paradigm shift is driven by landmark innovations, such as the generation of high-intensity, monoenergetic gamma-ray beams and novel particle beam formation techniques, which have revolutionized capabilities in nuclear materials inspection and medical applications[180]. Concurrently, breakthroughs in laser technology, now capable of generating intensities exceeding $10^{23}$ W/cm$^2$, have unlocked new experimental regimes in plasma physics and quantum electrodynamics[181]. These technological pillars underpin large-scale research infrastructures like the Extreme Light Infrastructure–Nuclear Physics (ELI-NP), which epitomizes the field's integrative potential by bridging laser physics, nuclear science, and materials research.

The exploration of photonic orbital angular momentum (OAM) represents a particularly innovative frontier within nuclear photonics, laying the groundwork for unprecedented quantum control. Early pioneering experiments demonstrated the feasibility of generating and utilizing OAM-carrying photons in regimes relevant to nuclear science. A pivotal milestone was the first direct experimental verification of photons carrying OAM at 99 eV, achieved in synchrotron radiation, which clearly demonstrated the characteristic spiral intensity distribution tied to radiation helicity[182]. Building on this foundation, subsequent research expanded into fundamental interactions, with studies on elastic photon-photon scattering revealing that OAM can significantly enhance the signal-to-noise ratio, a critical advantage for detecting notoriously weak quantum phenomena[183]. More recently, the potential of OAM has been theoretically extended to direct nuclear excitation, with proposals for using vortex beams to drive transitions in promising systems like the $^{229}$Th nuclear clock isomer[184]. Collectively, these crucial early steps established a new paradigm for manipulating quantum and nuclear systems using the unique properties of twisted light.

The generation and application of structured light are now pushing into the realm of nuclear physics and strong-field QED through the production of tailored high-energy photons. A primary mechanism is Inverse Compton Scattering (ICS), where a relativistic electron beam collides with an intense laser pulse, upscaling the photons to X-ray and γ-ray energies. The properties of the resulting high-energy radiation are directly imprinted by the structure of the laser pulse.

The transfer of OAM from a twisted laser photon to a γ-ray photon is a prime example of this capability[8,185]. While producing OAM light at optical wavelengths is routine, extending this to the MeV scale is extremely challenging but rich with applications. Vortex γ-ray beams carrying OAM can interact with atomic nuclei in novel ways, exciting giant multipole resonances[10] or enabling the control of precise nuclear spin polarization[186]. This opens the field of nuclear photonics, where structured light probes and manipulates nuclear states. Accelerator facilities in the 28-80 MeV electron beam energy range are pioneering this effort, aiming to generate OAM photons from 5 keV to over 100 keV and eventually into the MeV range[187]. The process involves finely tuning the collision between a relativistic electron beam and a structured laser pulse (e.g., a helical Laguerre-Gaussian or higher-order Bessel beam) that has been optimized via physics-aware training. The resulting vortex γ-rays could be used to selectively excite nuclear isomers, generate positrons with controlled angular momentum, or study laboratory astrophysics phenomena. Furthermore, the ability to generate quantum entangled pairs of photons in the MeV range using structured beams presents a new frontier for testing quantum foundations at high energies.

### 4.3. Quantum Electrodynamics and Quantum Optics of Strong Fields Physics

In addition to nuclear photonics and quantum electrodynamics (QED) at high frequency superintense EM fields, where structured X-rays and OAM transfer play a role, there exists another area of strong field physics, where quantum optics and QED start to play a central role: the realm of attophysics, where the laser frequencies are typically comparatively lower –

i.e.mid-IR – but they nevertheless generate high harmonics (HHG) up to the soft X-ray regime. Laser intensities are moderate while pulse durations are few-cycles short. Because the driving pulses are coherent, this becomes a powerful test-bed for structured light.

In its beginnings, the physics of intense laser-matter interactions was the physics of multiphoton processes. The theory was reduced then to high-order perturbation theory, while treating matter and light in a quantum manner. With the advent of chirped pulse amplification developed by D. Strickland and G. Mourou, which enabled generation of ultra-intense, ultra-short, coherent laser pulses, the need for a quantum electrodynamics description of EM fields practically ceased to exist and lost relevance[188]. Contemporary attoscience, and more generally ultrafast laser physics, awarded the Nobel Prize in 2023 to P. Agostini, F. Krausz, and A. L'Huillier, commonly uses the classical description of EM fields while keeping a fully quantum description of matter. The progress and successes of attoscience in the last 40 years have been spectacular, with an enormous amount of fascinating investigations in basic research and technology. Yet a central question remains: can ultrafast laser physics continue to advance without reintroducing QED and quantum optics into its description of light-matter interactions? Quantum optics has advanced towards the generation, control, and application of non-classical light states—such as Fock, squeezed, cat, and entangled states—that underpin emerging quantum technologies. However, these developments have largely been confined to low-photon-number and weak-field regimes, rather than high-intensity laser physics.

Recent pioneering works have demonstrated that conditioning intense-field processes—especially high-harmonic generation (HHG) and above-threshold ionization (ATI)—on measurable observables allows the generation of high-photon-number non-classical light. Optical Schrödinger cat states[189–193], multimode squeezed fields[192,194,195], and light-matter entangled states can now be engineered using strong-field processes, moving quantum optics firmly into the high-intensity regime. Moreover, bright squeezed states with intensity enough to drive or perturb HHG[196–198] and strong-field ionization[199] have been investigated. These developments have led to the emergence of quantum optics and QED of strong field processes, unifying ultrafast laser physics and quantum information science, captured in Fig. 6.

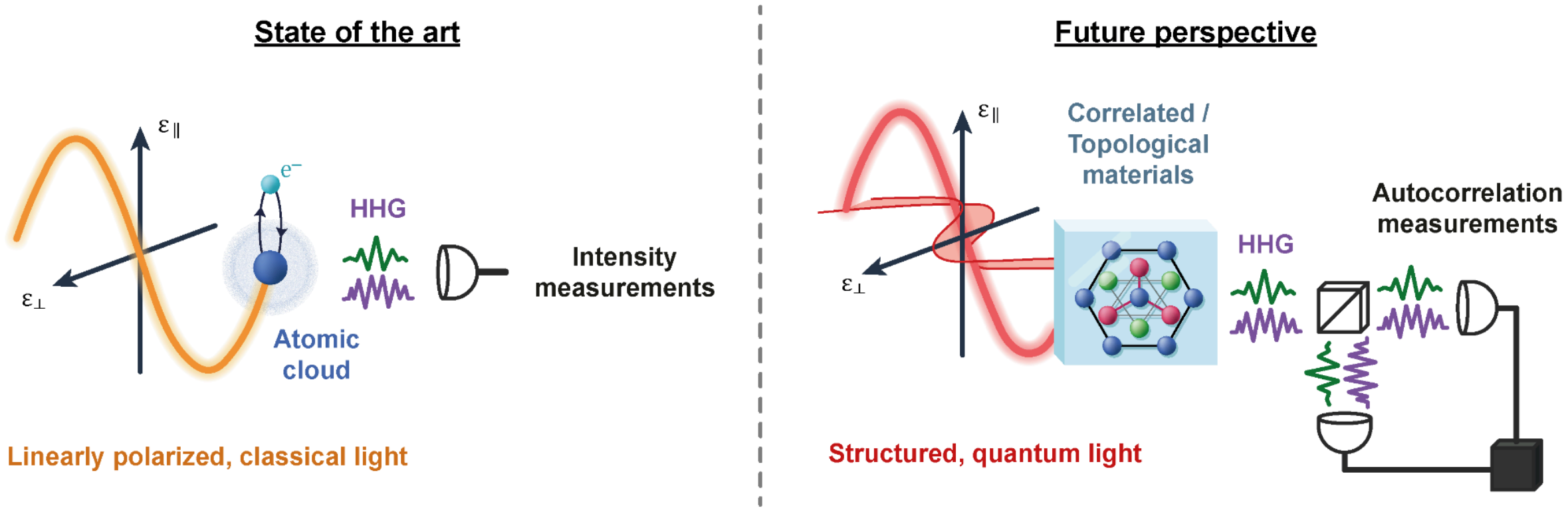

**Fig. 6 | Intersection of quantum optics and strong-field physics.** Possible future direction to go beyond the state of the art in the intersection between quantum optics and strong-field physics. The scenarios considered recently include the use of high-photon-number structured quantum light, or quantum laser fields interacting with correlated or topological materials, with their outputs characterized by autocorrelation measurement techniques.

There are three clear current and future challenges in this new area. They are all based on the necessary theoretical ingredient: development of quantum field-theoretical treatments of strong field physics[190].

- ***Generation of massively quantum states of light.*** New systematic methods should be established to produce large-photon-number entangled states, extending conditioning and post-selection techniques from atoms and molecules to correlated solids[191]. Generation of multimode squeezed and more exotic quantum states will be possible, going beyond the negligible depletion/excitation limit[194], or using HHG in resonant media[192]. Alternatively, one will use bright squeezed light[196,197], or its mixture with conventional intense laser pulses[198], to generate high harmonics. Structured light will be used to study geometrical, chiral, and topological states in QED of attoscience, potentially enabling to probe and control many-body quantum systems at an unprecedented scale[191].

- ***Generation of massively quantum states of light and matter.*** One will explore entanglement between quantized light fields and electronic states in complex materials, focusing on ultrafast processes such as ATI, HHG, and rescattering. This will provide the first systematic framework for observing and exploiting light-matter entangled states in solids, using similar approaches as mentioned above. Earlier work has shown that conditioning on ATI events or on distinct HHG recombination paths in molecules and simple solids can generate light-matter entanglement. Recent advances[200] concerning the reconstruction of the photoelectron density matrix provide a new testing ground. These studies have revealed how classical and quantum noise reduce purity, but the open question remains: how does photon entanglement influence the reconstructed density matrix, and can experimental data unambiguously signal the presence of underlying quantum correlations? This connects QED of attoscience to the broader framework of multi-fragment "Zerfall" processes. Again, structured light, carrying OAM, or spin, or both, maybe even presenting true topological knots in the polarization space, will play here c¡a crucial role.

- ***Characterization and exploitation of the generated states.*** Ultrafast quantum-optical methods to verify and quantify entanglement will need to be developed, providing new tools for nonlinear optics, precision metrology, and quantum information. This will open pathways to applying attosecond-scale entanglement across multiple fields. For the first steps toward applications in metrology and nonlinear optics, see [193,201]. Here, structured light will imply in this case novel ways of detection and applications.

### 4.4. High Energy Physics: Laser-Based Beam Collimation and Novel Particle Collisions

High Energy Physics requires ever increasing luminosity to explore subatomic interactions with increasing detail and precision. High luminosity colliders employ intense, high-energy beams which store extreme amounts of energy. The stored energy in the beam represents a hazard to machine operation. A collimation system is employed to protect vital components of the machine from lost beam particles, and for preventing high-amplitude "halo" particles from generating backgrounds in the detector. Traditional material-based collimation systems are struggling to meet the demands of high luminosity machines, as evidenced by struggles at facilities like SuperKEKB [202].

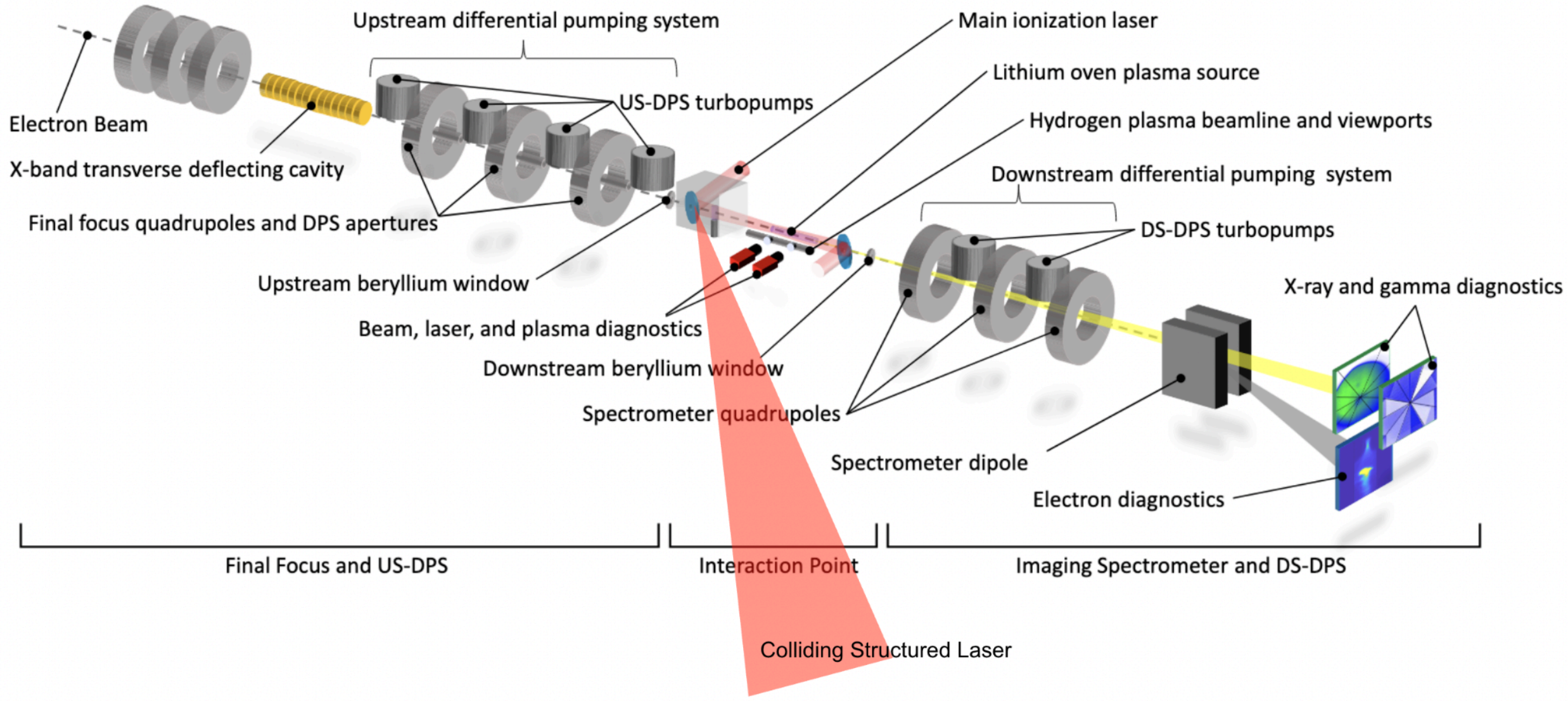

**Fig. 7 | Experimental schematic of the FACET-II Interaction Point (IP) Area.** A 10 GeV electron beam is focused to micron spot sizes and collides with a 10 TW-class Ti:S laser pulse. The e-beam and Compton backscattered photons are separated at the dipole. The suite of γ-ray diagnostics (right) measures the spatial profile and energy spectrum. Figure adapted from Ref. [203].

Structured intense laser beams offer a paradigm-shifting solution: a non-invasive "laser collimator." The concept leverages Compton scattering. A shaped laser pulse, such as a high-order Bessel beam with a dark central core, is aligned to interact only with the halo electrons [204]. The Compton interaction transfers momentum, kicking these unwanted particles onto lossy trajectories where they are safely absorbed at an energy aperture, while the core beam passes through the laser's null region unaffected. Additionally, laser-based systems for controlling the bunch intensity of beams in the FCC are currently being studied as a mechanism for avoiding the beam-beam flip-flop instability that arises from asymmetries in the colliding bunch intensity[205,206]. Experiments at FACET-II, such as the E320 and E344 programs represented in Fig. 7, are demonstrating this technology. They aim to show not only selective halo interaction but also the end-to-end optimization of γ-ray spectra for applications ranging from beam control to the production of novel particle states through γ-γ collisions[207,208].

This application perfectly demonstrates the need for the integrated toolkit:

- ***Light Shaping (Sec. 2):*** The laser profile must be specifically shaped (e.g., a "donut" mode) to interact only with the halo. Bessel beams are ideal due to their non-diffracting nature over long distances.
- ***Optimization (Sec. 3):*** The exact laser parameters (pulse duration, wavelength, profile) for optimal halo removal without disturbing the core are found using a digital twin of the accelerator and physics-aware training.
- ***Diagnostics & Feedback:*** The success of collimation is confirmed by measuring the spectrum and spatial profile of the Compton-scattered γ-rays using specialized spectrometers [209,210]. This measurement is fed back into the control loop for real-time optimization.

### 4.5 Next-Generation Electron Sources: Structured THz Fields and Miniaturized Waveguides

In advancing the frontier of structured light for particle acceleration, terahertz (THz)-based systems have emerged as a promising route toward compact, high-gradient electron accelerators. This section reviews key developments in THz-driven acceleration, categorized broadly into traveling-wave and advanced near-field schemes, highlighting how tailored mode structures and novel excitation mechanisms push the limits of energy gain and gradient.

**Traveling-Wave THz Acceleration.** A foundational architecture for THz acceleration adopts a traveling-wave configuration in dielectric-loaded waveguides (DLWs)[211]. Early work employed a circular DLW excited by single-cycle THz pulses generated via tilted-pulse-front optical rectification in lithium niobate. To couple energy efficiently, a segmented waveplate converted the linearly polarized input into a radially symmetric $TM_{01}$ mode, producing longitudinal electric fields for electron acceleration. Initial experiments delivering ~10 μJ of THz energy demonstrated a 7 keV energy gain over 3 mm[212], corresponding to an effective gradient of 2.5 MV/m—a milestone for THz-based linac-inspired designs.

Subsequent efforts sought higher acceleration gradients by increasing THz pulse energy beyond conventional crystal-based sources. One approach used coherent transition radiation from a relativistic electron bunch (460 fs, 850 pC, 30.4 MeV) incident on a metal foil, yielding 132 μJ THz pulses[213]. In a single DLW, this provided a 128 keV energy gain. By harnessing bidirectional THz emission, a cascaded setup with two aligned DLWs further increased the total energy gain to 204 keV, achieving a gradient exceeding 85 MV/m.

Extending the interaction length represents another critical pathway to higher net energy gain. In single-cycle regimes, velocity mismatch and group-velocity dispersion limit synchronous interaction, causing electrons to slip from the accelerating phase. Multi-cycle THz pulses mitigate this issue, preserving field integrity over longer distances despite a reduced peak field. For example, chirped-pulse beating in lithium nioribate produced 7 ps multi-cycle pulses of 2 μJ energy, enabling 10 keV energy gain in a rectangular DLW at a gradient of 2 MV/m[214].

**Evanescent-Wave and Plasmonic THz Acceleration.** To overcome limitations of free-space coupling and crystal damage thresholds, recent work has turned to evanescent THz fields and surface plasmon polaritons (SPPs). For instance, single-cycle evanescent waves at an internal reflection interface have supported gradients up to 6 MV/m, imparting 1.25 keV energy gain to 70 keV electrons while enabling attosecond streaking and temporal bunching[215].

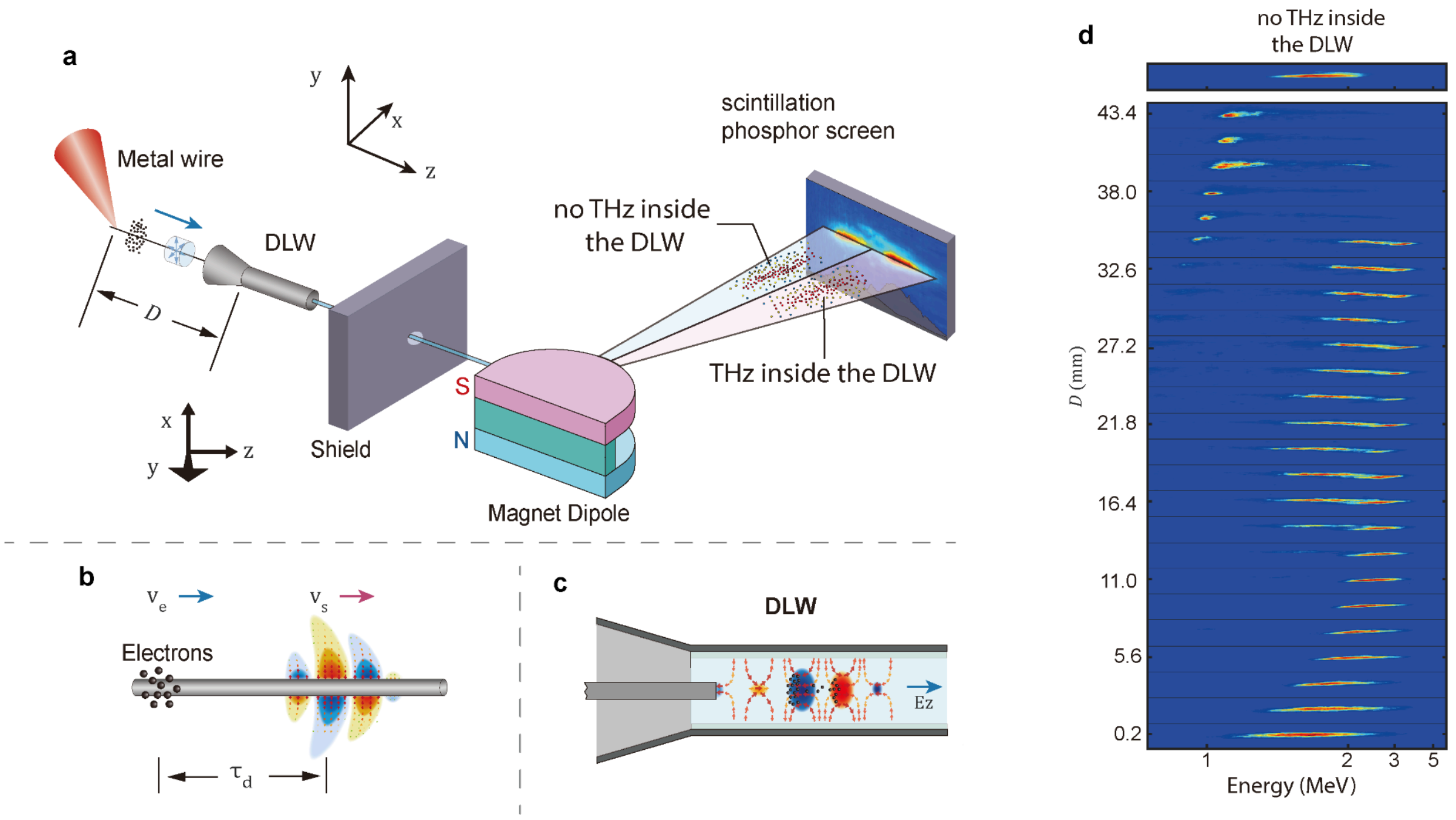

**Fig. 8 | Experimental setup of electron acceleration by THz surface waves.** A femtosecond laser pump pulse is focused onto the wire to excite the terahertz surface waves and the simultaneously produced electrons (a). Respective terahertz field profiles as the surface wave travels over the wire surface (b) and of the electromagnetic field emitted inside the DLW (c). Set

of experimental data showing the D-dependent electron energy modulations of the electron beam (d).

A particularly powerful approach integrates THz generation, amplification, and acceleration within a single waveguide structure. As shown in Fig. 8, using a thin metal wire, femtosecond laser excitation launches a THz SPP that co-propagates with ionized electrons[216]. Through coherent amplification, mJ-level THz SPPs can be confined to the wire—maintaining a $TM_{01}$-like mode ideal for acceleration. By concatenating the wire with a circular DLW, up to 85% coupling efficiency has been achieved. This scheme delivered a record 1.1 MeV energy gain over 5 mm, corresponding to a gradient of 210 MV/m. Future cascading through hollow metal tubes may further enable tabletop GeV-scale accelerators.

**Outlook and Challenges.** Despite remarkable progress—including demonstrated gradients beyond 200 MV/m in DLWs and multi-cell structures—several challenges persist. Scaling to practical systems requires THz pulse energies exceeding 10 mJ, currently limited by optical damage and coupling losses. Phase synchronization between electron bunches and THz fields remains critical for efficient interaction, necessitating micron-level fabrication precision and advanced beam control. Mitigating space-charge effects also demands spatiotemporally confined THz fields at the emission site.

Future advances will depend on: (1) new materials and waveguide designs tolerant to high intensities, (2) precision-engineered structures for extended interaction lengths, and (3) robust phase-stable THz sources. Success in these areas may not only enable compact MeV–GeV accelerators but also open new opportunities in ultrafast quantum control and light-source applications with unprecedented spatiotemporal resolution.

### 4.6. Structured THz Beams and Robust Optical Links

The unique properties of structured beams are driving transformative advancements in communications. Within the THz regime, structured beams provide a viable solution to overcome inherent limitations of conventional THz sources, including low output power, severe atmospheric attenuation, and vulnerability to physical blockages. Their engineered amplitude, phase, and polarization profiles underpin key capabilities, such as diffraction resistance, self-healing, and controlled bending, which are critical for establishing high-capacity, robust wireless links and directed energy systems.

Structured THz pulses, particularly those carrying OAM, can significantly advance communication system capacity and performance. Recent demonstrations have achieved data rates up to 112 Gbit/s using orbital-angular-momentum multiplexing and advanced modulation formats[217,218]. The long wavelengths of THz radiation make it susceptible to atmospheric absorption and scattering; however, the self-healing and diffraction-resistant properties of Bessel-type beams can enhance robustness and extend operational range in turbulent or scattering environments. Vectorial beam control mitigates signal degradation in free-space optical links caused by atmospheric turbulence. Tailored polarization and phase structures can maintain channel integrity, while the self-reconstructing nature of non-diffracting beams offers a significant advantage for maintaining links with moving platforms, such as satellites or unmanned vehicles, in contested environments.

**I. Generation and Control of Structured THz Beams**

A promising source of high-power structured THz waves is Dielectric Wakefield Acceleration (DWA). In this setup, a relativistic electron beam passes through a dielectric tube, emitting coherent Cherenkov radiation[219]. The radiation traveling in this structure can be intrinsically imprinted with OAM by the photonic design of the dielectric liner or after emission using spiral phase plates or metasurfaces developed for high-power applications. This approach provides unequaled peak power and tunable, narrow bandwidth, exploiting the new degrees of freedom permitted by OAM to create secure, high-bandwidth communication channels resistant to jamming and eavesdropping.

Alternatively, metasurfaces, composed of subwavelength artificial atoms, enable precise control over the phase, amplitude, and polarization of THz waves. A notable strategy leverages the nonlinear Pancharatnam-Berry (PB) phase: by selectively exciting anisotropic meta-atoms using a femtosecond laser, researchers have achieved direct generation of vortex beams (carrying orbital angular momentum, OAM) across the 0.8–1.4 THz band. This approach integrates wavefront shaping directly into the beam generation process, bypassing the bandwidth constraints of conventional linear metasurfaces and offering an integrated solution for 6G communications and high-resolution radar[220,221].

Another critical advancement lies in programmable liquid-crystal-based metasurfaces. One representative architecture adopts a cross-switch layout to generate row-column encoded patterns, enabling high-directivity, two-dimensional beam steering on a 56×56 element array. This design drastically simplifies the feed network for large-scale arrays and demonstrates robust beamforming/scanning capabilities in the W-band, laying a hardware foundation for integrated sensing-communication systems in the THz range[222].

**II. Robust Communications with Structured THz Beams**

The intrinsic properties of structured beams offer dual solutions to mitigate THz signal degradation in complex environments: enhancing communication capacity via spatial multiplexing and improving link resilience through the beams' inherent physical characteristics.

In terms of capacity, OAM multiplexing represents a transformative approach by exploiting orthogonal spatial modes in the THz regime. Distinct OAM modes, characterized by unique topological charges, establish independent data channels in spatially multiplexed links, directly enabling high-capacity wireless communication[223]. In terms of resilience, THz Bessel-type and Airy beams exhibit remarkable self-reconstruction capabilities, allowing them to recover their intensity profiles after encountering obstacles[224]. This property is pivotal for maintaining non-line-of-sight (NLOS) links in challenging scenarios; for example, recent studies have integrated physics-informed learning frameworks to dynamically optimize Airy beam parameters (e.g., cubic phase, focal length) in real time, further enhancing link stability. Additionally, the robustness of structured beams (e.g., steady optical vortex beams, STOVB) in turbulent environments is validated by a significant reduction in temporal intensity fluctuations[225].

**III. Challenges and Future Perspectives**

Despite substantial progress, several bottlenecks remain for translating these technologies into practical systems:

- **Power and Efficiency**: The conversion efficiency of many passive/active metasurfaces remains low (e.g., nonlinear PB metasurfaces down to ~$10^{-8}$), limiting practical link budgets and energy transfer rates. A promising pathway to overcome this lies in the synergistic co-design of state-of-the-art, high-power THz sources—such as those based on optical rectification with tilted-pulse fronts[226,227], relativistic electron beams[228,229], or laser-plasma interactions[230]—with advanced adaptive and static THz optics. By integrating these high-flux sources with efficient, robust beam-shaping elements like metasurfaces or dielectric wakefield structures, the field can transcend the limitations of individual components. This integrated approach will be crucial for achieving the high effective isotropic radiated power (EIRP) necessary for long-range, high-bandwidth links and efficient directed energy applications, pushing structured THz systems toward practical deployment.

- **System Integration**: Integrating dynamic beam-control devices, high-power THz sources, and real-time feedback algorithms into compact, low-power-consumption platforms is a critical next step.
- **Channel Modeling**: Accurate channel models that account for the propagation of twisted, self-accelerating, and non-diffracting beams in complex, turbulent atmospheres are still under development.

Future advancements will likely focus on novel high-power THz source architectures, more efficient and faster programmable metasurfaces, and deep integration of artificial intelligence for real-time adaptive beam control. Success in these areas will firmly establish structured THz beams as the foundation for next-generation wireless communications and directed energy systems, redefining the boundaries of transmission range, link robustness, and communication capacity.

## 5. FUTURE PERSPECTIVES AND GRAND CHALLENGES

The integrated vision of structured light control, AI-driven optimization, and groundbreaking applications, as outlined in the previous sections, charts an ambitious course for the next decade of high-field laser-matter physics. However, transitioning from pioneering laboratory demonstrations to robust, standardized, and widespread technologies presents a set of interconnected grand challenges. Addressing these bottlenecks is paramount to ushering in the next revolution in photonics and accelerator science.

### 5.1. Material Science for Extreme Optics: The Search for New Nonlinear and High-Damage-Threshold Materials

The generation and control of structured light at extreme intensities are fundamentally constrained by material properties. The optical elements described in Section 2—whether static gratings, metasurfaces, or plasma modulators—push against current physical limits.

- **Damage Threshold:** The foremost challenge is optical damage. As laser systems continue to scale in peak and average power (e.g., towards multi-PW and high-repetition-rate systems), the demand for optical components that can withstand immense electromagnetic fields grows more acute. While progress has been made with materials like KDP for polarization control[53] and plasma-based components that are inherently damage-resistant, there is a critical need for new material platforms. Research must focus on wide-bandgap semiconductors, novel ceramic composites, and nano-engineered coatings that offer orders-of-magnitude improvements in laser-induced damage threshold (LIDT).
- **Nonlinearity and Dispersion Engineering:** Advanced pulse shaping techniques, particularly those operating in the UV and THz regimes, rely on nonlinear optical processes. The efficiency and fidelity of these processes are dictated by the nonlinear coefficients and dispersion properties of the materials. Future progress depends on the development of materials with giant, tailorable nonlinearities and engineered dispersion profiles. This could involve meta-materials, periodically poled crystals with novel domain structures, or gases and plasmas with optimized nonlinear optical responses.
- **Active Optical Materials:** The ultimate goal of dynamic, high-speed control requires materials that can actively modulate light properties at picosecond or faster timescales without sacrificing damage resistance. This could involve exploring novel electro-optic, acousto-optic, or all-optical switching mechanisms in new material systems that can operate at high peak powers and high repetition rates simultaneously.

### 5.2. Closing the Loop in Real-Time: The Integration of AI, Diagnostics, and Adaptive Control at MHz Rates

The optimization engine in Section 3 is powerful but typically runs offline. A promising next step is real-time, closed-loop control at the native laser rate (kHz–MHz). Early demonstrations of model-free policy gradient (PG) methods established that optical systems could be directly optimized from measurements—no digital twin required. In these implementations, a parameterized policy was iteratively updated in the direction of the measured reward gradient, enabling *in situ* learning for tasks such as wavefront correction and optical classification[96,231] To further enhance the speed of convergence, Proximal Policy Optimization (PPO) has been introduced as a stable and data-efficient method[232]. By reusing measurements for multiple gradient steps and enforcing clipped updates, PPO constrains each policy update within a trust region, improving convergence robustness. This approach achieves rapid and resilient optimization across diverse structured-light applications, including energy focusing, holographic image synthesis, in-situ aberration correction, all-optical classification[233]. By reducing measurements per update and tolerating noise, misalignment, and drift, this hardware-in-the-loop learning forms a practical basis for autonomous control of an experimental set-up.

Building on this capability, realizing a true MHz-rate loop, three elements must be co-designed: fast data reduction, fast actuation, and a lightweight controller at the edge.

- **Data Acquisition and Processing**. Diagnostics in Sec. 2.5 yield dense wavefront, spectral, and temporal data each shot—a microsecond-scale bottleneck. The in-situ PPO strategy alleviates this by (i) deriving compact reward signals directly from measured intensities (e.g., focus ratio, PSNR, class score), (ii) reusing a single batch for multiple gradient steps, and (iii) enforcing stable updates via clipping. In practice, wavefront cameras and spectro-temporal sensors feed FPGA/ASIC/DSP pipelines for low-latency feature extraction and RL updates, while a supervisory host adjusts hyperparameters over aggregated shots. This converts raw diagnostics into actionable feedback within microseconds, preparing the control signal for the actuator stage.
- **High-Speed Actuation**. With low-latency feedback available, the remaining constraint is modulator response time. While SLMs offer flexibility, learned phase and amplitude control laws can communicate with faster hardware—integrated electro-optic modulator arrays, resonant EO phase shifters, or acousto-optic programmable dispersive filters (AOPDFs) with higher bandwidth. The same reward definitions used in training drive shot-synchronous updates; the policy maps diagnostics to actuator set-points at line rate, and a slower outer loop refines the policy over many shots. This completes the inner loop on the actuation path.
- **AI at the Edge**. To coordinate these layers and maintain robustness, a hierarchical architecture balances accuracy and latency. An offline digital twin (when available) seeds policies and operating regions, while an online edge controller—a compact network implementing PPO updates or a distilled surrogate—applies per-shot corrections at hardware speed. Continuous in-situ adaptation narrows the simulation–reality gap by tracking unmodeled aberrations, thermal drift, and target changes. Together, streamed diagnostics, fast actuation, and edge AI close the loop in real time at kHz–MHz repetition rates, with periodic supervisory re-optimization to sustain performance under evolving conditions.

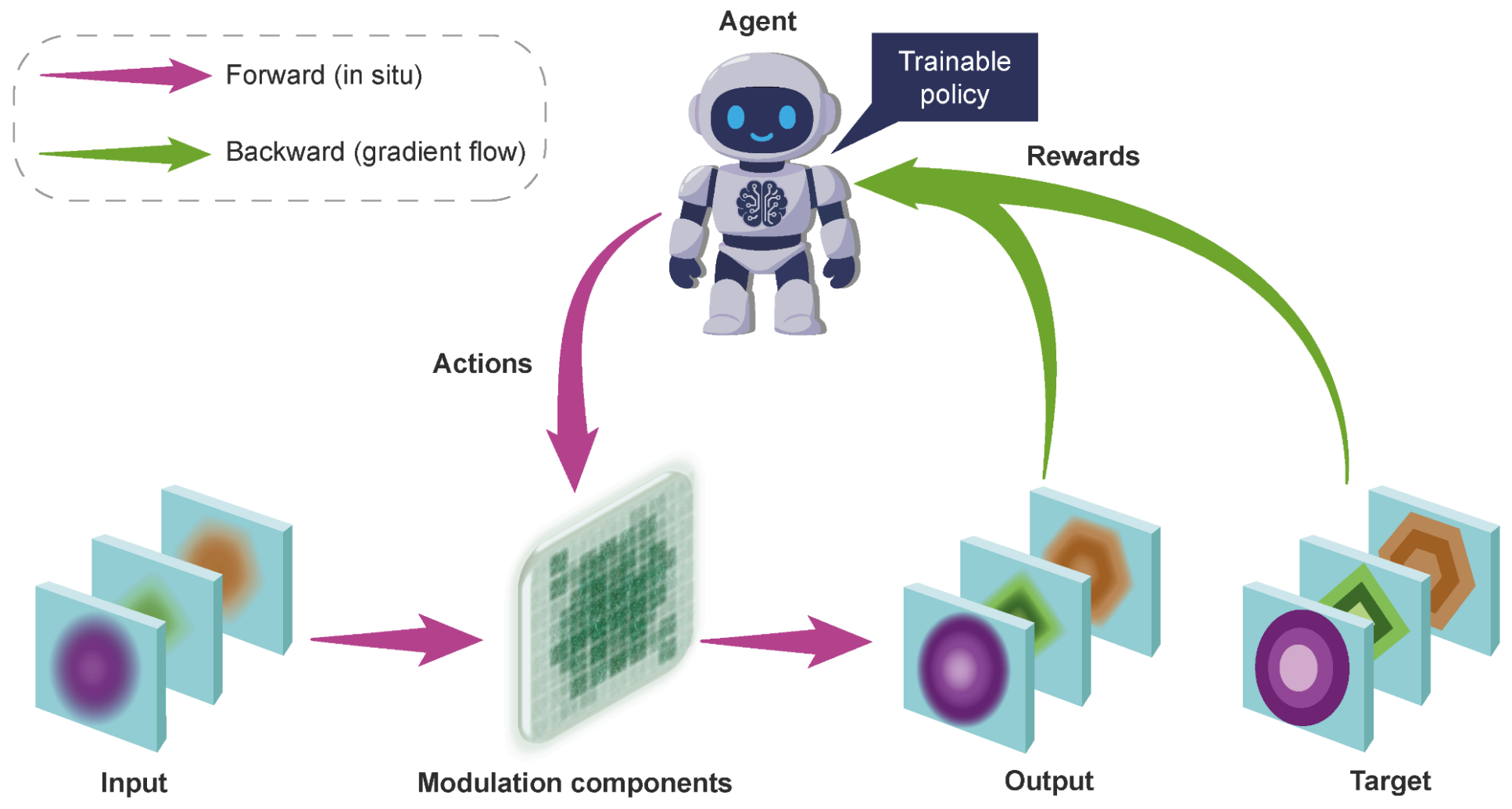

**Fig. 9 | Schematic of a high-speed, closed-loop control system for real-time optimization of structured light.** The system integrates three co-designed elements: (1) High-Speed Diagnostics (e.g., a wavefront sensor and spectrometer) that provide single-shot, low-latency feature extraction (e.g., focus ratio, PSNR). (2) An Edge AI Controller that processes this data and executes a lightweight optimization algorithm (e.g., Proximal Policy Optimization) to compute control updates at kHz–MHz rates. (3) High-Speed Actuators (e.g., electro-optic modulator arrays or AOPDFs) that apply the optimized phase/amplitude patterns to the laser pulse. This closed loop enables autonomous, real-time correction and optimization of structured light fields under evolving experimental conditions.

### 5.3. From Macro to Quantum: Applying Structured Light Control to Quantum Systems and Single-Particle Interactions

Thus far, the focus has been on controlling collective, classical phenomena. A profound frontier is applying these same principles to control quantum systems and engineer light-matter interactions at the single-particle level.

- **Quantum Electron Light Sources:** The concept of "free-electron quantum optics" is emerging, where the quantum state of electrons is prepared and manipulated to generate quantum light[234]. Shaping the wavefunction of individual electron pulses using structured light fields could enable the generation of heralded single X-ray photons, entangled photon pairs, or squeezed light at hard X-ray wavelengths, opening new avenues for quantum sensing and quantum information science with high-energy photons.

- **Control of Quantum Materials:** The ability to craft intense, arbitrary electromagnetic waveforms could be used to coherently control the quantum state of matter. This goes beyond simple excitation; it involves using the precise temporal and spatial structure of light to steer chemical reactions, manipulate topological phases in materials, or prepare exotic quantum states in solids that are otherwise inaccessible.
- **Strong-Field QED and Vacuum Polarization:** At the highest intensities, the structured laser field itself could be used to probe high-field classical and quantum electrodynamic effects. We have already shown that the unique properties of flying-focus beams can be exploited to enhance the effects of the self electromagnetic field of an electron, known as radiation reaction[235,236]. The longer interaction time as compared to a fixed focus beam, indeed, allows one to perform high-field experiments under more controlled conditions. Analogously we have shown that vacuum birefringence can also be measured in principle exploiting the same property of flying-focus beams [235–237], by combining the latter with already available coherent sources of x-rays. Finally, structured light featuring definite orbital angular momentum can be also employed for transporting ultrarelativistic electron and positron beams over much longer distances than conventional fixed-focus beams[238]. Tailored light is essential for these experiments to separate the tiny QED signals from overwhelming background processes.
- On a different aspect, we have pointed out how structured light can be an extremely useful tool to shape electron wave packets. If such techniques can be extended to the ultrarelativistic domain, they will provide unique possibilities for example, to produce coherent radiation at high frequencies, where the shape of the electron's wave packet will play an important role. Needless to say, also the high-frequency radiation emitted by such appropriately shaped electron wave packets will have different properties in terms of polarization and frequency content as compared to standard non-linear Compton scattering, which can trigger further applications.

### 5.4. Standardization and Community Adoption: Developing Benchmarks and Open-Source Tools for the Field

For this field to mature and move beyond a few specialized laboratories, it must overcome the challenge of standardization.

- **Benchmarking and Datasets:** There is a critical need for standardized benchmark problems, simulated and experimental datasets, and well-defined metrics of success. This would allow different research groups to compare the performance of their shaping algorithms, digital twins, and control systems objectively. Community-wide challenges, similar to those in computer vision or machine learning, could accelerate progress.
- **Open-Source Software and Hardware:** The development of open-source software frameworks for building digital twins (e.g., extensions of LUME[91]), optimizing light-matter interactions, and controlling experiments is essential. Similarly, open-hardware designs for key components, such as plasma cell designs for PLMs or diagnostic setups, would lower the barrier to entry and foster collaboration.
- **Interoperability:** As systems become more complex, ensuring that different components—simulation codes, control software, hardware interfaces—can work together seamlessly is a major challenge. Adopting common data standards and communication protocols will be key to building the integrated, automated laboratories of the future.

### 5.5. A Call to Action: Key Research Initiatives for the Next Decade

To overcome these challenges and capitalize on the opportunities, a coordinated effort is required. Key research initiatives should focus on:

1. **A Dedicated Materials Discovery Program:** A large-scale initiative to discover, synthesize, and characterize new optical materials specifically designed for high-power, structured light applications, leveraging high-throughput computation and experimental screening.
2. **Integrated Testbed Facilities:** Establishing dedicated user facilities that combine state-of-the-art high-power lasers, advanced beam shaping tools, and real-time AI control systems. These testbeds would serve as incubators for developing and benchmarking new technologies and algorithms.
3. **The "Moore's Law" for Photonic Control:** A concerted effort to drive down the latency and cost and drive up the speed and capacity of optical modulation and diagnostic systems, akin to the evolution of integrated circuits.
4. **Theory and Simulation for Inverse Design:** Investing in the development of new multi-scale, multi-physics simulation codes that are inherently differentiable and designed from the ground up to be integrated with AI training loops.

5. **Complex topologically structured light-matter interaction:** Beyond conventional vortices and vector beams, the recently emerging complex topological quasiparticle structured light[239–241], e.g. optical skyrmions, hopfions, toroidal vortices, etc., possess intricate 3D polarization, phase, and energy-flow topologies, offering new degrees of freedom to tailor strong light–matter interactions at ultrafast timescales[242].

## 6. CONCLUSION

The era of brute-force, unstructured light in high-field physics is drawing to a close. As we have outlined, a profound transformation is underway, moving us toward a new paradigm of intelligent photonics—a synergistic convergence of advanced light shaping, AI-driven inverse design, and real-time adaptive control. This transition is not merely incremental; it represents a fundamental shift from observing light-matter interactions to commanding them with unprecedented precision.

The journey from foundational concepts to groundbreaking applications is built upon three pillars. First, the electromagnetic toolkit has evolved beyond conventional spatial light modulators to include robust static optics and, most promisingly, programmable plasma-based modulators, enabling full vectorial control of light at extreme intensities. Second, the optimization engine of physics-informed digital twins and physics-aware training provides the algorithmic foundation to navigate the staggering complexity of these systems, solving the inverse problem and automating discovery. Finally, these capabilities are already enabling breakthroughs across diverse frontiers, from programmable particle accelerators and nuclear photonics with structured γ-rays to robust communications and non-invasive beam collimation for future colliders.

Realizing the full potential of this vision hinges on the community's ability to tackle the overarching grand challenges: the search for new materials capable of withstanding extreme fields, the integration of AI and diagnostics for real-time control at MHz rates, and the extension of these principles to the quantum realm. This endeavor is inherently interdisciplinary, demanding collaboration between optical scientists, material engineers, computer scientists, and physicists.

The path forward is clear. By continuing to forge these connections and confront these challenges, we can usher in a future where light is not just a tool but a programmable force, unlocking new chapters in fundamental science and empowering technologies that are today beyond our imagination.

**Funding Acknowledgements.** U.S. Department of Energy contracts DE-AC02-76SF00515 and DE-SC0022559, U.S. Air Force Office of Scientific Research contract FA9550-23-1-0409, U.S. National Science Foundation contracts 2431903 and 2431903, U.S. Office of Naval Research contract N00014-24-1-2038, National Key Research and Development Program of China (2022YFA1604400), and National Natural Science Foundation of China (U23A6002), Singapore Ministry of Education (MOE) AcRF Tier 1 grants (RG157/23 & RT11/23), Singapore Agency for Science, Technology and Research (A*STAR) MTC Individual Research Grants (M24N7c0080).